\documentclass[preprint,aps,showpacs]{revtex4}
\usepackage{graphicx}
\usepackage{amsmath}

\begin{document}

\title{Nonlinear oscillations and waves in multi-species cold plasmas}

\author{Prabal Singh Verma\footnote{prabal.singh.verma@ipp.mpg.de}}
\affiliation{Technische Universit\"at Berlin, ER 3-2, Hardenbergstr. 36a, 10623 
Berlin, Germany, and Max-Planck/Princeton Center for Plasma Physics}

\date{\today}

\begin{abstract}
%It has recently been shown using a standard perturbative analysis that the  
%electrostatic oscillations in an electron-positron-ion plasma exhibit phase mixing even in the linear regime. 
%As a result, oscillations break at an arbitrarily small amplitude in such plasmas. 
%In this work 
The spatio-temporal evolution of nonlinear oscillations in multi-species 
plasma is revisited to provide more insight into the physics of phase mixing by constructing 
two sets of nonlinear solutions up to the second order. The first solution exhibits perfect oscillations in the linear regime 
and phase mixing appears only nonlinearly in the second order as a response to the ponderomotive forces.  
This response can be both direct and indirect. The indirect contribution of the ponderomotive forces 
appears through self-consistently generated low frequency fields. 
Furthermore, the direct and 
indirect contributions of the ponderomotive forces on the phase mixing process is explored and it is 
found that the indirect contribution is negligible in an electron-ion plasma 
and it disappears in the case of electron-positron plasma, yet represents an equal contribution 
in the electron-positron-ion plasma.
However, the second solution 
does not exhibit any phase mixing due to the absence of ponderomotive forces but results in an undistorted nonlinear traveling wave.   
These investigations have relevance for laboratory/astrophysical multi-species plasma. 

\end{abstract}


\maketitle
\section{Introduction}

Phase mixing is a phenomenon that causes plasma oscillations/waves to break at an amplitude much 
smaller than the critical amplitude \cite{dawson,infeld-1989,kaw-1973,nappi}. Wave breaking is a process that destroys the coherent oscillations 
and thus limits the amplitude of the wave \cite{psv-pre-2012,prabal-2011,davidson,davidson-book}. The concept of wave breaking is being used 
extensively, {\it e.g.} in particle acceleration 
\cite{taji-daw,modena,malka-2002,malka-2006,malka-2009,psv-prl-2012,ion-acc-2015,ele-acc-2015}, 
fast ignition \cite{tabak} and plasma heating \cite{pukhov}.
Wave breaking via phase mixing arises when the plasma frequency for some 
physical reason acquires a spatial dependence \cite{sudip-2011}. 
As a result, plasma species oscillate at different local frequencies at different positions in space 
and lead to trajectory crossing (and thus wave breaking) after a finite time.
{Thus phase mixing is a slow process that leads to wave breaking. Slow wave breaking of plasma oscillations is being studied extensively 
in various physical regimes \cite{new1,new2,new3,new4}  }.

The phase mixing process has been interpreted as mode coupling where the initial energy loaded on the 
fundamental mode goes nonlinearly into the higher and higher harmonics with time and 
damps the primary mode \cite{kaw-1973,sudip-1999}.
In the relativistic case the plasma frequency picks up the spatial dependence due to 
the relativistic mass variation effect \cite{psv-prl-2012,infeld-1989-rel,sudip-2009,sudip-2011,chandan-prl-2013}. 
However, in a cold homogeneous electron-ion (e-i) plasma, the frequency of the oscillations obtains its spatial dependency 
due to the modification of the ion background in response to the 
self-consistently generated slow (DC) terms \cite{sudip-1999}. {These slow terms do not have time dependence therefore we call them DC terms.}
These DC terms may arise either linearly as an response to the zero 
frequency mode \cite{psv-pop-2011} and/or nonlinearly in response to the ponderomotive forces \cite{sudip-1999,sudip-2011}. 
In the previous work it has been shown that, although the plasma oscillations in an 
arbitrary mass ratio cold plasma phase mix and break at an arbitrarily small amplitude, there exists a nonlinear traveling
wave solution in such e-i plasmas which does not exhibit phase mixing \cite{psv-pop-2011}.  

Physics of the multi-species (electron-positron-ion) plasmas can be more complex than of e-i plasmas. These 
plasmas exist naturally in various astrophysical environments \cite{astro1,astro2,astro3,astro4,astro5} and are also produced in laboratory by laser matter 
interaction \cite{lab1}. An enormous amount of effort has been put in understanding nonlinear wave phenomena in 
such plasmas in various physical regimes \cite{epi1,epi2,epi3,epi4,epi5,epi6,epi7,epi8,epi9,epi10}.  
In a recent publication the space time evolution of cold electron-positron-ion (e-p-i) plasma has been investigated using a  
perturbative method and it has been shown that plasma oscillations experience phase mixing even in the linear regime 
and break at an arbitrarily low amplitude \cite{chandan-pop-2014}. 
In order to provide more insight into the physics of phase mixing, the space-time evolution of e-p-i plasma is revisited and 
it is demonstrated that there exist two sets of solutions in e-p-i plasma, one displays phase mixing only 
nonlinearly in the second order solution and the other does not encounter any phase mixing but results in 
a nonlinear traveling wave 

It is well known that in electron-ion plasmas the background species (ions) respond to ponderomotive 
forces either directly or 
indirectly through low frequency self-consistent fields and hence attain inhomogeneity in space \cite{sudip-1999}. 
In this work we analyze the direct and indirect contributions of ponderomotive forces on 
the phase mixing of multi-species plasmas and find that in an electron-ion plasma the effect of  
indirect contributions is negligible and it vanishes in the case of electron-positron plasma. However, 
both the effects exhibit equal contributions on the phase mixing of nonlinear oscillations in an electron-positron-ion plasma.
In order to do this analysis, the effect of the zero frequency mode on 
the nonlinear oscillations has been ignored by the choice of the 
initial conditions which satisfy the linear dispersion relation of the system. 
Of course, the initial conditions can be chosen arbitrarily. However, the initial conditions as chosen in 
this work allow us to categorize the cause of phase mixing 
during the nonlinear evolution. This issue is later discussed in detail by considering various cases.

The paper is organized as follows. Section II contains the basic equations governing the 
dynamics of the multi-species plasma and its space-time evolution. Section III provides a nonlinear solution which shows 
pure oscillations at the linear level and exhibits phase mixing in the second order. In section IV, nonlinear 
solution is constructed which does not show any phase mixing.
Section V contains the summary and discussion of all the results presented in this manuscript. 

\section{Governing equations and perturbation analysis}
The basic equations describing the motions of a cold multi-species plasma are 
the continuity equations, momentum equations and the Poisson equation,  
\begin{equation}\label{econ}
 \frac{\partial n_{e}}{\partial t}+\frac{\partial(n_{e}v_{e})}{\partial x}=0
\end{equation}
\begin{equation}\label{icon}
 \frac{\partial n_{p}}{\partial t}+\frac{\partial(n_{p}v_{p})}{\partial x}=0
\end{equation}
\begin{equation}\label{emom}
\frac{\partial v_e}{\partial t} + v_e\frac{\partial v_e}{\partial x}= -e E/m_e
\end{equation}
\begin{equation}\label{imom}
\frac{\partial v_p}{\partial t} + v_p\frac{\partial v_p}{\partial x}=  e E/m_p
\end{equation}
 \begin{equation}\label{poi}
  \frac{\partial E}{\partial x}= 4 \pi e (n_{i0} + n_p - n_e)
 \end{equation}
Here ions are assumed to be static and $n_{i0}$ is the equilibrium background ion density. The subscript `$p$' stands for positive ions (which are 
lighter than the background ions) and `$e$' stands for the electrons. The remaining symbols have their usual meanings {and {\it cgs} unit is used throughout}.
The plasma is quasi-neutral {\it i.e.}, $n_{i0} + n_{p0} = n_{e0}$, where $n_{p0}$ and $n_{e0}$ are equilibrium densities of 
positive ions (mobile) and electrons respectively. Let us define, 
 $n_p = n_{p0} + \delta n_p$ and $n_e = n_{e0} + \delta n_e$ and introduce a new variable $\delta n_d$ such that 
$\delta n_d = \delta n_p - \delta n_e = n_{i0} + n_p - n_e$. 

In the perturbative analysis the nonlinear solution for $\delta n_{d}$ can be expressed as,

$$\delta n_{d} = \delta n_{d}^{(1)} + \delta n_{d}^{(2)} + \delta n_{d}^{(3)} + ...$$
Other physical quantities can be described in a similar fashion.

%%%%%%%%%%%%%%%%%%%%%%%%%%%%%%%%%%%%%%%%%%%%%%%%%%%%%%%%%%%%%%%%%%%%%%%%%%%%%%%%%%%%%%%%%%%%%%%
\subsection{First order solution}
%%%%%%%%%%%%%%%%%%%%%%%%%%%%%%%%%%%%%%%%%%%%%%%%%%%%%%%%%%%%%%%%%%%%%%%%%%%%%%%%%%%%%%%%%%%%%%%

The set of Eqs.\eqref{econ}-\eqref{poi} in the first order approximation can be combined to give,
 \begin{equation}\label{pdend1}
 \frac{\partial^2 \delta n_{d}^{(1)}}{\partial t^2}+\omega_{p}^2\delta n_{d}^{(1)}=0,
\end{equation}
 where,  $$\omega_{p}^2 = \frac{4 \pi n_{0e} e^2}{m_e} +\frac{4 \pi n_{0p} e^2}{m_p} = \omega_{pe}^2 + \omega_{pp}^2.$$ 
The quantities $\omega_{pe}$, $\omega_{pp}$ are   
electron and ion plasma frequencies respectively. The solution of Eq.\eqref{pdend1} can be expressed as, 
  \begin{equation}\label{nd11}
  \delta n_{d}^{(1)} = A(x) \cos \omega_{p}t + B(x) \sin \omega_{p}t
  \end{equation}
  Here $A$ and $B$ are constants which are to be determined from the initial conditions.

%%%%%%%%%%%%%%%%%%%%%%%%%%%%%%%%%%%%%%%%%%%%%%%%%%%%%%%%%%%%%%%%%%%%%%%%%%%%%%%%%%%%%%%%%%%%%%%
\subsection{Second order solution}
%%%%%%%%%%%%%%%%%%%%%%%%%%%%%%%%%%%%%%%%%%%%%%%%%%%%%%%%%%%%%%%%%%%%%%%%%%%%%%%%%%%%%%%%%%%%%%%
Again, the set of Eqs.\eqref{econ}-\eqref{poi} in the second order approximation can be expressed as,

\begin{equation}\label{econ1}
 \frac{\partial \delta n_{e}^{(2)}}{\partial t}+n_{e0}\frac{\partial(v_{e}^{(2)})}{\partial x}
+\frac{\partial(\delta n_{e}^{(1)}v_{e}^{(1)})}{\partial x}=0,
\end{equation}
\begin{equation}\label{icon1}
 \frac{\partial \delta n_{p}^{(2)}}{\partial t}+n_{p0}\frac{\partial(v_{p}^{(2)})}{\partial x}
+\frac{\partial(\delta n_{p}^{(1)}v_{p}^{(1)})}{\partial x}=0,
\end{equation}
\begin{equation}\label{emom1}
\frac{\partial v_e^{(2)}}{\partial t} + v_e^{(1)}\frac{\partial v_e^{(1)}}{\partial x}= -e E^{(2)}/m_e,
\end{equation}
\begin{equation}\label{imom1}
\frac{\partial v_p^{(2)}}{\partial t} + v_p^{(1)}\frac{\partial v_p^{(1)}}{\partial x}=  e E^{(2)}/m_p,
\end{equation}
 \begin{equation}\label{poi1}
  \frac{\partial E^{(2)}}{\partial x}= \delta n_{d}^{(2)},
 \end{equation}
 
These equations can be combined as follow,

 \begin{eqnarray}\label{nd2}
\nonumber \frac{\partial^2 \delta n_{d}^{(2)}}{\partial t^2}+ \omega_p^2 \delta n_d^{(2)} = -\frac{\partial^2}{\partial x^2}
  \Big[\frac{n_{e0} v_e^{(1)2} - n_{p0} v_p^{(1)2}}{2}\Big] \\ - \frac{\partial^2}{\partial x \partial t}
\Big[{\delta n_{p}^{(1)} v_p^{(1)} + \delta n_{e}^{(1)} v_e^{(1)}}\Big]
\end{eqnarray}

%%%%%%%%%%%%%%%%%%%%%%%%%%%%%%%%%%%
\section{Nonlinear solution (I)}
%%%%%%%%%%%%%%%%%%%%%%%%%%%%%%%%%%%

 Let us choose the first set of initial conditions as, 
  \begin{eqnarray}
 \nonumber \delta n_e = n_{e0} \delta \cos(kx), \delta n_p = n_{p0} \Delta \delta \cos(kx), \\
  v_e = 0, v_p = 0, \text{ \hspace{0.1cm} where\hspace{0.1cm} }  \Delta = \frac{m_e}{m_p}.%\hspace{0.4cm}   
  \end{eqnarray}
  {Here $\delta$ is the maximum amplitude of the normalized electron density perturbation $|\delta n_e /n_{e0}|$. }  
  These initial conditions lead to the first order solution as, 
   \begin{equation}\label{solnd11}
   \delta n_{e}^{(1)} = n_{e0} \delta \cos (kx) \cos (\omega_{p}t)
  \end{equation}
  \begin{equation}\label{solv11}
   \delta v_{e}^{(1)} = \delta \frac{\omega_p}{k} \sin (kx) \sin (\omega_{p}t)
  \end{equation}
  \begin{equation}\label{solV11}
   \delta n_{p}^{(1)} = -n_{p0} \delta \Delta \cos (kx) \cos (\omega_{p}t)
  \end{equation}
  \begin{equation}\label{solns11}
   \delta v_{p}^{(1)} = -\delta \frac{\omega_p}{k} \Delta \sin (kx) \sin (\omega_{p}t)
  \end{equation}
    \begin{equation}\label{solE11}
  E^{(1)} = -4\pi e (n_{e0}+n_{p0} \Delta) \frac{\delta}{k} \sin (kx) \cos (\omega_p t)
  \end{equation}
  The set of Eqs.\eqref{solnd11}-\eqref{solE11} exhibits a pure oscillatory solution in the first order and 
there is no DC term present in any physical quantity. Thus, the phase mixing is impossible on the linear 
level. However, in \cite{chandan-pop-2014} the signature of phase mixing is seen even in the linear solution {which appears in 
response to the zero frequency mode of the system \cite{psv-pop-2011,sudip-1999}}.  
 
We can now write down the second order solution  for $\delta n_{d}^{(2)}$ and $E^{(2)}$ as follow,
  \begin{eqnarray}\label{solnd2}
   \nonumber \delta n_{d}^{(2)} =  \frac{\delta^2}{2}(n_{e0}-\Delta^2 n_{p0}) \cos (2kx) \\ \Big[2\cos (\omega_p t)-(1+\cos (2\omega_{p}t))\Big]
  \end{eqnarray}
    \begin{eqnarray}\label{solE2}
  \nonumber   E^{(2)} = \frac{\delta^2}{4} \frac{4\pi e(n_{e0}-\Delta^2 n_{p0})}{k}\sin (2kx) \\ \Big[2\cos (\omega_p t)-(1+\cos (2\omega_{p}t))\Big]
  \end{eqnarray}

We notice that the density $\delta n_{d}^{(2)}$ and hence electric field $ E^{(2)}$ have obtained the DC terms. These DC terms are 
arising due to the imbalance of the ponderomotive forces on both the species. 

In order to figure out the contribution of the nonlinearly generated DC electric field, which is   
an indirect effect of the ponderomotive forces, on the dynamics of the 
electron fluid let us write down Eq.\eqref{emom1} by retaining only time averaged 
terms as, 

\begin{eqnarray}\label{emom2}
\nonumber \frac{\partial v_e^{(2)}}{\partial t} + \frac{\delta^2}{4k}{(\omega_{pe}^2 + \omega_{pp}^2)} \sin(2 kx) \\ = 
\frac{\delta^2}{4k} {(\omega_{pe}^2 - \omega_{pp}^2 \Delta)} \sin (2kx),
\end{eqnarray}

Let us now consider four cases to categorize the cause of phase mixing: (1) electron plasma oscillations where ions are static, (2) electron-ion (e-i) oscillations where ions are mobile, 
(3) electron-positron(e-p) oscillations (4) electron-positron-ion (e-p-i) oscillations where ions are static.

{\it Electron plasma oscillation --} Eq.\eqref{emom2} reduces to electron plasma oscillations case if we substitute $\Delta,\omega_{pp} = 0$ {\it i.e.},  

\begin{eqnarray}\label{emom3}
 \frac{\partial v_e^{(2)}}{\partial t} + \frac{\delta^2}{4k}{(\omega_{pe}^2)} \sin(2 kx)  = 
\frac{\delta^2}{4k} {(\omega_{pe}^2)} \sin (2kx),
\end{eqnarray}

Thus, from Eq.\eqref{emom3} we notice that even in the electron plasma oscillations $E^{(2)}$ contains a DC term, however, the effect of this 
DC field on the electrons gets cancelled by the ponderomotive forces. Therefore electron plasma oscillations do not exhibit any phase 
mixing. 

{\it Electron-ion oscillation --} Again in Eq.\eqref{emom2} if we substitute $\omega_{pp}^2 = \omega_{pe}^2 \Delta$, it reduces to the e-i oscillations case 
and we get the following equation,

\begin{eqnarray}\label{emom4}
\nonumber \frac{\partial v_e^{(2)}}{\partial t} + \frac{\delta^2}{4k}{(\omega_{pe}^2 + \omega_{pe}^2 \Delta)} \sin(2 kx) \\ = 
\frac{\delta^2}{4k} {(\omega_{pe}^2 - \omega_{pe}^2 \Delta^2)} \sin (2kx),
\end{eqnarray}

From Eq.\eqref{emom4} we can see that $E^{(2)}$ has a contribution of the order of $\sim \Delta^2$ on the phase mixing of plasma oscillations 
which is an indirect contribution of the ponderomotive forces. However, 
the ponderomotive force itself gives a $\sim \Delta$  contribution which is a direct contribution. 
Thus, the contribution of $E^{(2)}$ to the phase mixing for 
$\Delta << 1$ is negligible and Eq.\eqref{emom4} in this limit gives, 

\begin{eqnarray}\label{emom44}
 v_e^{(2)} \sim t {\delta^2}{\Delta} \sin(2 kx).  
\end{eqnarray} 

Using Eq.\eqref{emom44} in Eq.\eqref{econ1} we get, 

\begin{eqnarray}\label{econ44}
 \delta n_e^{(2)} \sim t^2 {\delta^2}{\Delta} \cos(2 kx).  
\end{eqnarray} 

This shows a clear signature of phase mixing which is also consistent with equation(13) in ref.\cite{sudip-1999} in the limit $\Delta << 1$.

{\it Electron-positron oscillation --} This case can be obtained from Eq.\eqref{emom2} by substituting $\omega_{pp} = \omega_{pe}$ and $\Delta = 1$ 
which gives, 

\begin{eqnarray}\label{emom5}
\nonumber \frac{\partial v_e^{(2)}}{\partial t} + \frac{\delta^2}{4k}{(\omega_{pe}^2 + \omega_{pe}^2)} \sin(2 kx) \\ = 
\frac{\delta^2}{4k} {(\omega_{pe}^2 - \omega_{pe}^2)} \sin (2kx),
\end{eqnarray}

From Eq.\eqref{emom5} we notice that the R.H.S. vanishes. This indicates that $E^{(2)}$ in this case does not contribute to the phase mixing, {\it i.e}, 
the indirect effect of the ponderomotive forces on the phase mixing of plasma oscillations is absent. However, 
the ponderomotive force directly plays the main role in the 
phase mixing process. 
This can be confirmed by substituting $\Delta = 1$ in equations (27) and (37) 
of ref. \cite{psv-pop-2011} which give $E^{(2)} = 0$. We further confirm this fact by substituting $\Delta = 1$ in equation(12) in ref.\cite{sudip-1999}, 
which indicates zero contribution of $E^{(2)}$ to the phase mixing.   

{\it Electron-positron-ion oscillation --} We can obtain this case by substituting $\Delta = 1$ in Eq.\eqref{emom2} such that, 

\begin{eqnarray}\label{emom6}
\nonumber \frac{\partial v_e^{(2)}}{\partial t} + \frac{\delta^2}{4k}{(\omega_{pe}^2 + \omega_{pp}^2)} \sin(2 kx) \\ = 
\frac{\delta^2}{4k} {(\omega_{pe}^2 - \omega_{pp}^2)} \sin (2kx),
\end{eqnarray}
From Eq.\eqref{emom6} we notice that e-p-i oscillations have a special feature that both the DC forces are contributing equally in the phase mixing process. The Eq.\eqref{emom6} can be further solved to give,

\begin{eqnarray}\label{emom66}
 v_e^{(2)} \sim t \frac{\delta^2}{2k} \omega_{pp}^2 \sin(2 kx).  
\end{eqnarray} 

Using Eq.\eqref{emom66} in Eq.\eqref{econ1} we get,

\begin{eqnarray}\label{econ66}
 \delta n_e^{(2)} \sim n_{e0} \frac{t^2}{2} {\delta^2}\omega_{pp}^2 \cos(2 kx).  
\end{eqnarray}

{As is already discussed that phase mixing will occur when frequency of the system acquires a spatial dependence. Thus if we write down the equation of motion up to third order \cite{sudip-1999,chandan-pop-2014} one can clearly see frequency of the system acquiring a 
spatio-temporal dependence} and hence Eq.\eqref{econ66} confirms the initiation of phase mixing and wave breaking \cite{sudip-1999,psv-pop-2011}.
We can now note here that, equation (23) in ref. \cite{chandan-pop-2014} gives a different expression for $\delta n_e^{(2)}$ which is,

\begin{eqnarray}\label{chandan_ne}
 \delta n_e^{(2)} \sim n_{e0} \frac{t^2}{2} {\delta^2}\omega_{pp}^2 (1-\frac{\omega_{pp}^2}{\omega_{p}^2})^2  \cos(2 kx).  
\end{eqnarray}

In Eq.\eqref{chandan_ne} we see an additional coefficient $(1-\frac{\omega_{pp}^2}{\omega_{p}^2})^2$ in the 
expression of $\delta n_e^{(2)}$. This  
might be arising as a result of the DC term in the linear solution and ultimately appearing into the 
scaling of the phase mixing \cite{chandan-pop-2014}. However, we observe that ${\omega_{pp}^2}<{\omega_{p}^2}$, 
therefore if we expand $(1-\frac{\omega_{pp}^2}{\omega_{p}^2})^2$ and retain only the dominant term, Eq.\eqref{chandan_ne} reduces to Eq.\eqref{econ66}. Thus, we see that the zero frequency mode plays a negligible role when 
compared to the ponderomotive forces in the nonlinear evolution of the system.  
Now, from Eq.\eqref{econ66} we can obtain the scaling of phase mixing time for the nonlinear oscillations in the 
e-p-i plasmas as, $t_{mix} \sim \frac{2}{{\delta} \omega_{pp}}$. This scaling is similar to the one found by 
Kaw et al. \cite{kaw-1973} with 
inhomogeneous static ion background. From this scaling we learn that for a finite value of $\delta$ as the equilibrium density of 
positron decreases {\it i.e,} when positron are replaced by static
ions the phase mixing time of the oscillations increases and as $n_{p0} \rightarrow 0$, $t_{mix} \rightarrow \infty$.

%%%%%%%%%%%%%%%%%%%%%%%%%%%%%%%%%%%
\section{Nonlinear solution (II)}
%%%%%%%%%%%%%%%%%%%%%%%%%%%%%%%%%%% 
In this section, we construct a nonlinear solution in e-p-i plasmas which does not exhibit phase mixing. 
In order to do that we choose the following initial conditions, 

  \begin{eqnarray}
 \nonumber \delta n_e^{(1)} = n_{e0} \delta \cos(kx), \delta n_p^{(1)} = n_{p0}  \delta \cos(kx), \\ 
  v_e^{(1)} = \frac{\omega_{p}} {k} \delta \cos(kx), v_p^{(1)} = -\frac{\omega_{p}} {k}  \delta \cos(kx),% \\ 
  \end{eqnarray}
  these initial conditions lead to the first order solution as, 
   \begin{equation}\label{solnd_tw}
   \delta n_{e}^{(1)} = n_{e0} \delta \cos (kx - \omega_{p}t). 
  \end{equation}
  \begin{equation}\label{solv_tw}
   \delta v_{e}^{(1)} = \delta \frac{\omega_p}{k} \cos (kx - \omega_{p}t). 
  \end{equation}
  \begin{equation}\label{solV_tw}
   \delta n_{p}^{(1)} = -n_{p0} \delta  \cos (kx - \omega_{p}t). 
  \end{equation}
  \begin{equation}\label{solns_tw}
   \delta v_{p}^{(1)} = -\delta \frac{\omega_p}{k}  \cos (kx - \omega_{p}t). 
  \end{equation}
    \begin{equation}\label{solE_tw}
  E^{(1)} = -4\pi e (n_{e0}+n_{p0} ) \frac{\delta}{k} \sin (kx - \omega_{p}t). 
  \end{equation}
  The set of Eqs.\eqref{solnd_tw}-\eqref{solE_tw} exhibits a pure traveling wave solution in the first order. 
Using this set of equations in Eq.\eqref{nd2},

\begin{equation}\label{nd22_tw}
 \frac{\partial^2 \delta n_{d}^{(2)}}{\partial t^2}+ \omega_p^2 \delta n_d^{(2)} = 3 \delta^2 \omega_p^2 (n_{e0}-n_{p0}) \cos(2kx-2\omega_p t).
\end{equation}
Now the solution of Eq.\eqref{nd22_tw} can be expressed as,
\begin{eqnarray}\label{solnd222_tw}
 \nonumber \delta n_{d}^{(2)} = C(x) \cos \omega_p t + D(x) \sin \omega_p t \\
 - \delta^2 (n_{e0}-n_{p0}) \cos(2kx-2\omega_p t).
\end{eqnarray}
 Let us now choose the second order perturbations such that $C(x) = D(x) = 0$. Thus Eq.\eqref{nd22_tw}
 becomes,

 \begin{equation}\label{solnd2222_tw}
 \delta n_{d}^{(2)} =  - \delta^2 (n_{e0}-n_{p0}) \cos(2kx-2\omega_p t).
 \end{equation}
  Now from Eq.\eqref{poi1} second order solution for the electric field can be written as, 
  \begin{equation}\label{solE2222_tw}
  E^{(2)} =  - 4 \pi e \frac{\delta^2}{2k} (n_{e0}-n_{p0}) \sin(2kx-2\omega_p t).
 \end{equation}  

Similarly other quantities can be obtained as,
 \begin{equation}\label{solve2222_tw}
  v_{e}^{(2)} =   \frac{\delta^2}{2k} \frac{\omega_{pe}^2}{\omega_p} \cos(2kx-2\omega_p t).
 \end{equation}

\begin{equation}\label{solne2222_tw}
  \delta n_{e}^{(2)} =   \frac{\delta^2}{2} n_{e0}(1 + \frac{\omega_{pe}^2}{\omega_p^2}) \cos(2kx-2\omega_p t).
 \end{equation}

 \begin{equation}\label{solvp2222_tw}
  v_{p}^{(2)} =   \frac{\delta^2}{2k} \frac{\omega_{pp}^2}{\omega_p} \cos(2kx-2\omega_p t).
 \end{equation}

\begin{equation}\label{solnp2222_tw}
  \delta n_{p}^{(2)} =   \frac{\delta^2}{2} n_{e0}(1 + \frac{\omega_{pp}^2}{\omega_p^2}) \cos(2kx-2\omega_p t).
 \end{equation}

The set of Eqs.\eqref{solE2222_tw}-\eqref{solnp2222_tw} exhibits a pure traveling wave solution. Since this 
non-linear solution does not have any DC term we do not expect any phase mixing. 

{Because travelling waves get excited 
when the laser pulse propagates through the plasma and the phase velocity of these waves is decided by 
the group velocity of the laser pulse \cite{gibbon}, one can in principle excite the travelling waves 
(Eqs.\eqref{solE2222_tw}-\eqref{solnp2222_tw}) by choosing a laser pulse whose group velocity matches the 
phase velocity ($\omega_p/k$ in this case) of the waves in multispecies plasma (Eqs.\eqref{solE2222_tw}-\eqref{solnp2222_tw}). 
Although, these travelling waves are the solution of zero measure like relativistic plasma waves \cite{psv-prl-2012,AP-1956} and 
would phase mix if perturbed slightly, they can be well detected experimentally because phase mixing is a slow process. }

 \section{Summary} 
  In this paper, it has been demonstrated that although the oscillations in e-p-i plasmas 
may exhibit phase mixing at the linear level, there exist two sets of nonlinear solutions which do not show
 this feature. 
  One exhibits perfect oscillations at the linear level and displays phase mixing only non-linearly in the second 
  order as a consequence of both direct and indirect responses to the ponderomotive forces. The direct and indirect effect 
of the ponderomotive forces on the phase mixing of 
nonlinear oscillations in a multi-species plasma have further been examined and it is discovered that 
the indirect contribution of the ponderomotive forces is insignificant in an electron-ion plasma 
and it dies out in case of the electron-positron plasma. However, both direct and indirect effects exhibit equal 
contributions in case of electron-positron-ion plasma.   
It is also found that in multi-species plasmas the zero frequency mode plays an ignorable role in the phase 
mixing process during the nonlinear evolution of the system when compared with ponderomotive forces.
 On the contrary, the second 
 solution does not show any phase mixing due to non-appearance of ponderomotive forces and gives rise to an exact traveling wave. 
These traveling 
waves are solutions of zero measure {like relativistic plasma waves \cite{psv-prl-2012}}. Therefore, they will also phase mix when perturbed slightly. 
Nevertheless, these waves may be detected experimentally since phase mixing is a slow process.
These studies have relevance in laser-matter interaction experiments as well as in astrophysical environments.

\section{Acknowledgement}  The author would like to thank Tapan Chandra Adhyapak for the discussion and 
useful suggestions and Wolf-Christian M\"uller for carefully reading the manuscript and giving valuable 
comments. This work is supported by Max-Planck/Princeton Center for Plasma Physics.
%The author is also grateful to Wolf-Christian M\"uller for his careful reading of the 
%manuscript and for suggesting valuable English corrections . 

\end{document}